\documentclass{mem}
\usepackage{natbib}\usepackage{txfonts}\usepackage{balance}
\usepackage{graphicx}
\usepackage[a4paper,breaklinks,dvipdfm]{hyperref}
\idline{000}{001}
\begin{document}
\def\teff{$T\rm_{eff }$}
\def\kms{$\mathrm {km s}^{-1}$}

\title{Constraining fundamental physics with combined variations of fundamental constants}

   \subtitle{}

\author{
A.M.R.V.L. \,Monteiro\inst{1,2,4} 
\and M.C. \,Ferreira\inst{1,2}
\and M.D. \,Juli\~ao\inst{1,3}
\and C.J.A.P. \,Martins\inst{1}
          }

  \offprints{A.M.R.V.L. \,Monteiro}
 
 \institute{
Centro de Astrof\'isica, Universidade do Porto, 
Rua das Estrelas, 4150-762 
Porto, Portugal
\and
Faculdade de Ci\^encias, Universidade do Porto, 
Rua do Campo Alegre, 4150-007 
Porto, Portugal
\and
Faculdade de Engenharia, Universidade do Porto, 
Rua Dr Roberto Frias, 4200-465 
Porto, Portugal
\and
Department of Applied Physics, Delft University of Technology, 
P.O. Box 5046, 2600 GA 
Delft, The Netherlands
\email{mmonteiro@fc.up.pt}
}

\authorrunning{Monteiro {\it et al.}}

\titlerunning{Constraining Fundamental Physics with Varying Constants}

\abstract{
We discuss how existing astrophysical measurements of various combinations of the fine-structure constant $\alpha$, the proton-to-electron mass ratio $\mu$ and the proton gyromagnetic ratio g$_{p}$ towards the radio source PKS1413$+$135 can be used to individually constrain each of these fundamental couplings. While the accuracy of the available measurements is not yet sufficient to test the spatial dipole scenario discussed in this workshop (and elsewhere in this volume), our analysis serves as a proof of concept as new observational facilities will soon allow significantly more robust tests. Importantly, these measurements can also be used to obtain constraints on certain classes of unification scenarios, and we compare the constraints obtained for PKS1413$+$135 with those previously obtained from local atomic clock measurements (and discussed in the previous contribution).
}

\maketitle{}

\section{Introduction}

Nature is characterised by a set of physical laws and fundamental dimensionless couplings, which have always been assumed to be space time-invariant. For the former this is a cornerstone of the scientific method: it is hard to imagine how one could do science at all if it were not the case. For latter, however, this is not the case: our current working definition of a fundamental constant is simply any parameter whose value cannot be calculated within a given theory, which must therefore be found experimentally. Very little is known about these couplings, their role in physical theories or even to what extent they are really fundamental.

Fundamental couplings are known to run with energy, and in many extensions of the standard model they will also roll in time and ramble in space. As a consequence, calculations valid for our tangible space-time might not be correct for different times or places. A detection of varying fundamental couplings will be revolutionary: it will automatically prove that the Einstein Equivalence Principle is violated (and therefore that gravity can not be purely geometry), and that there is a fifth force of nature. If so different regions of the universe will effectively have different physical phenomenology.

Any Grand Unified Theory predicts a specific relation between variations of $\alpha$ and $\mu$, and this relation will be highly model-dependent. However, this is a blessing rather than a curse, as it implies that simultaneous measurements of both can provide us with key consistency tests of the underlying physical mechanisms. 

The present work if focused on three measurements of different combinations of the fine-structure constant $\alpha$, the proton-to-electron mass ratio $\mu$ and the proton gyromagnetic ratio $g_{p}$ towards the radio source PKS1413$+$135 at redshift $z  \approx 0.247$ \citep{murphy,darling,kanekar}. Together, these allow us to individually constrain each of these couplings. Although these constraints are relatively weak, forthcoming observational facilities will significantly improve existing measurements. Thus, our analysis is also a proof of concept, since improved measurements will allow for much stronger tests.

In addition to their intrinsic test as precision consistency tests of the standard cosmological model, these tests of the stability of fundamental constants can also be used to obtain constraints on certain classes of unification scenarios. This has been previously done for local (redshift $z = 0$) using comparisons of atomic clocks \citep{ferreira1}, as reported in M. Juli\~ao's contribution to these proceedings. In this sense the present work is an extension of this formalism to the early universe. Further details can be found in \citet{ferreira2}.

\begin{table*}
\caption{\footnotesize
Current combined measurements (with their one-sigma uncertainties) at $z \approx 0.247$ towards the radio source PKS1413$+$135.}
\label{pks}
\begin{center}
\begin{tabular}{lccc}
\hline
$Q_{AB}$ & $\Delta Q_{AB}/Q_{AB}$ & Reference \\
\hline
\\
$\alpha^{2} g_{p}$ & $(-2.0 \pm 4.4) \times 10^{-6}$ & \citet{murphy}\\
$\alpha^{2\times 1.57} g_{p} \mu^{1.57}$ & $(5.1 \pm 12.6) \times 10^{-6}$ & \citet{darling} \\
$\alpha^{2\times 1.85} g_{p} \mu^{1.85}$ & $(-11.8 \pm 4.6) \times 10^{-6}$ & \citet{kanekar} \\
\\
\hline
\end{tabular}
\end{center}
\end{table*}

\section{Quasar absorption spectra}

The spectrum of the radio source PKS1413$+$135 includes a number of interesting molecular absorption as well as emission lines. From comparisons of different lines one can obtain measurements of several combinations of the fundamental couplings $\alpha$, $\mu$ and $g_{p}$. Specifically, we will consider the best available measurements from each of three different (and independent) techniques, which are summarised in Table \ref{pks}. One should note that the first two are null results while the last measurement is a detection at more than two standard deviations. In all cases we define relative variations as

\begin{equation}
\frac{\Delta Q}{Q} = \frac{Q(z=0.247) - Q(z=0)}{Q(z=0}.
\end{equation}

\begin{figure}[t!]
\resizebox{\hsize}{!}{\includegraphics[width=0.5\textwidth]{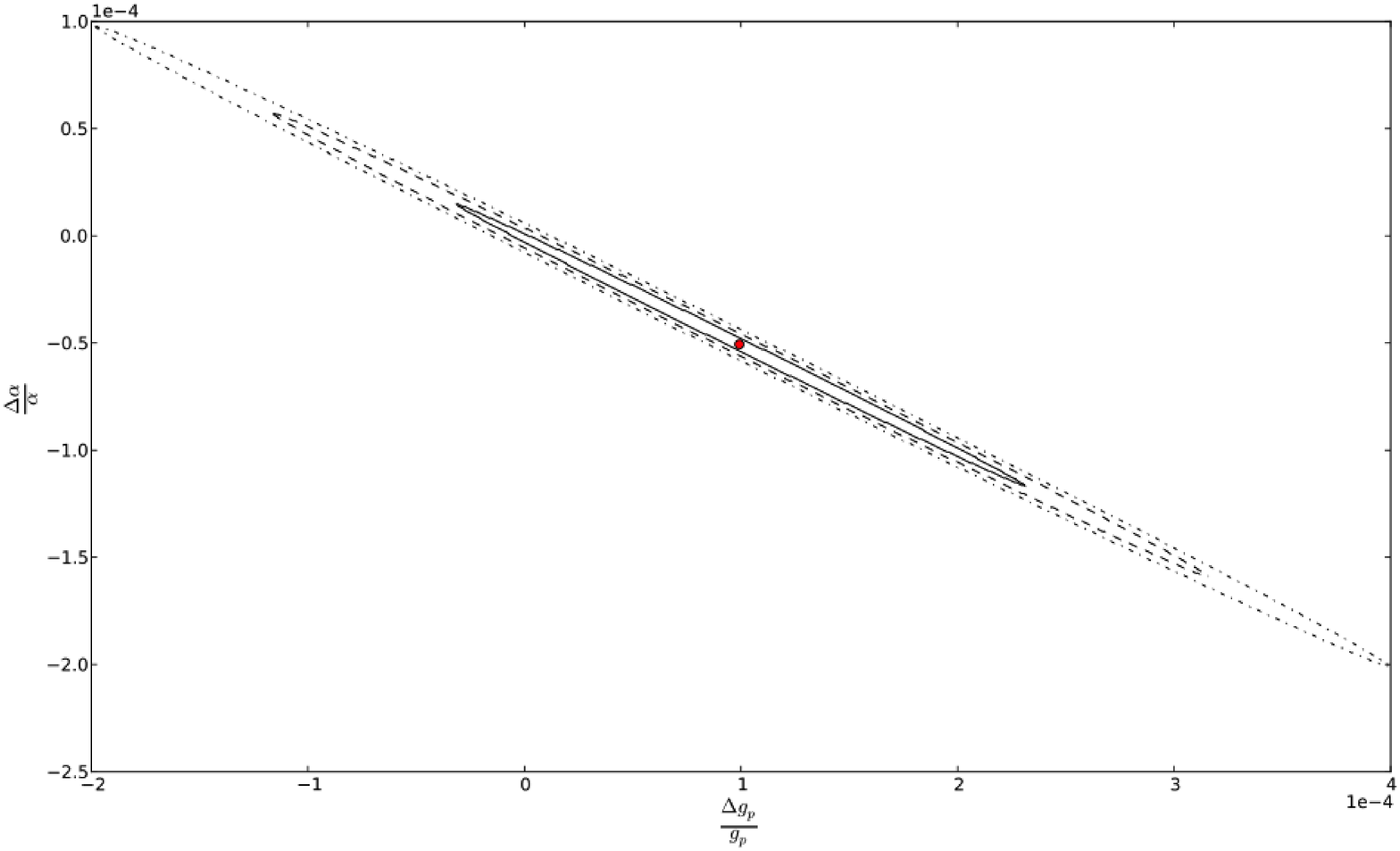}}
\resizebox{\hsize}{!}{\includegraphics[width=0.5\textwidth]{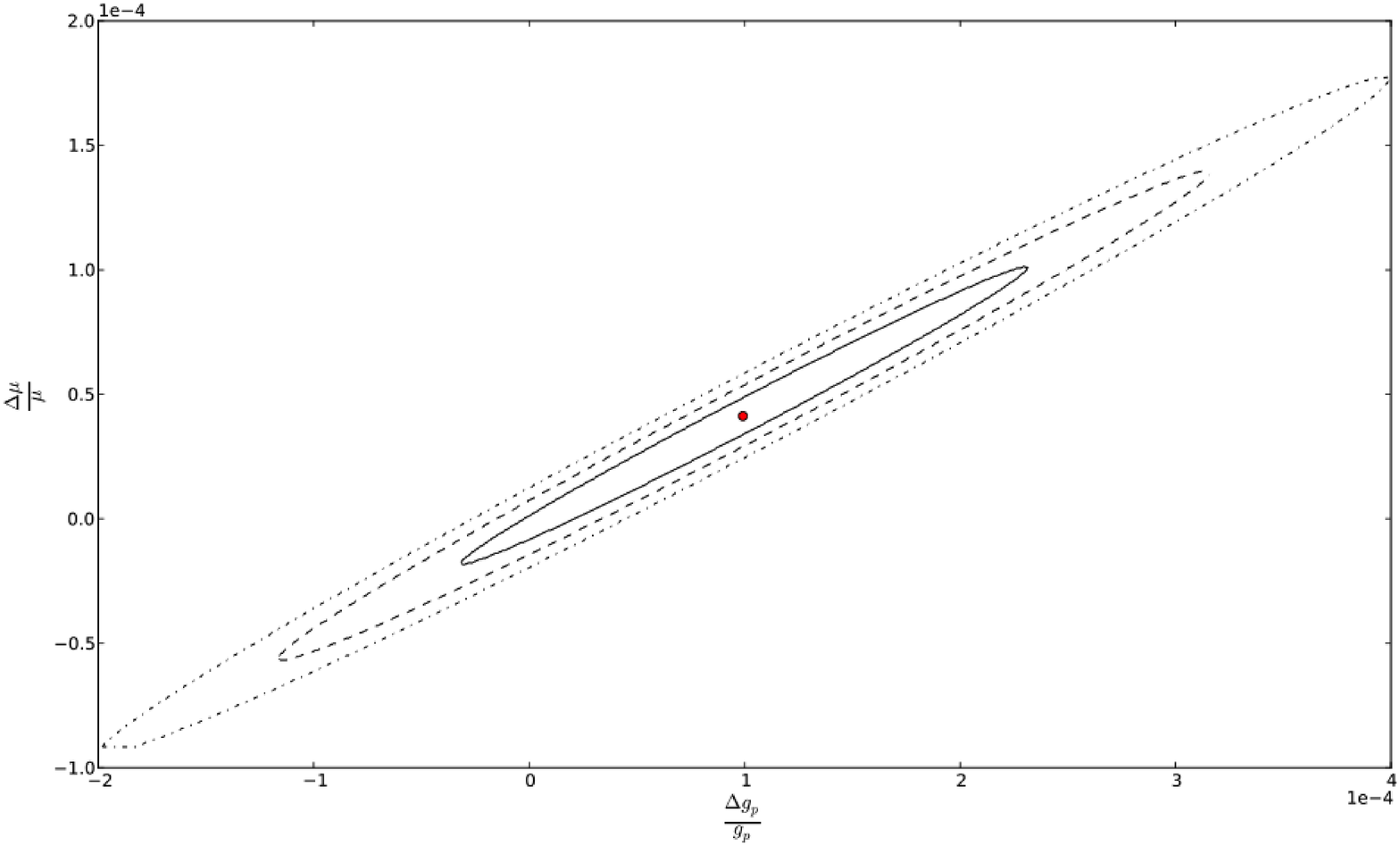}}
\caption{\footnotesize
Two-dimensional likelihood for the relative variations of $\alpha$, $\mu$ and $g_{p}$, between $z\approx0.247$ and $z=0$. Solid, dashed and dotted lines correspond to one-, two- and three-sigma contours (68.3\%, 95.4\% and 99.97\% likelihood, respectively). Reprinted, with permission, from \citet{ferreira2}.
}
\label{qua}
\end{figure}

\begin{figure}[t!]
\resizebox{\hsize}{!}{\includegraphics[width=0.5\textwidth]{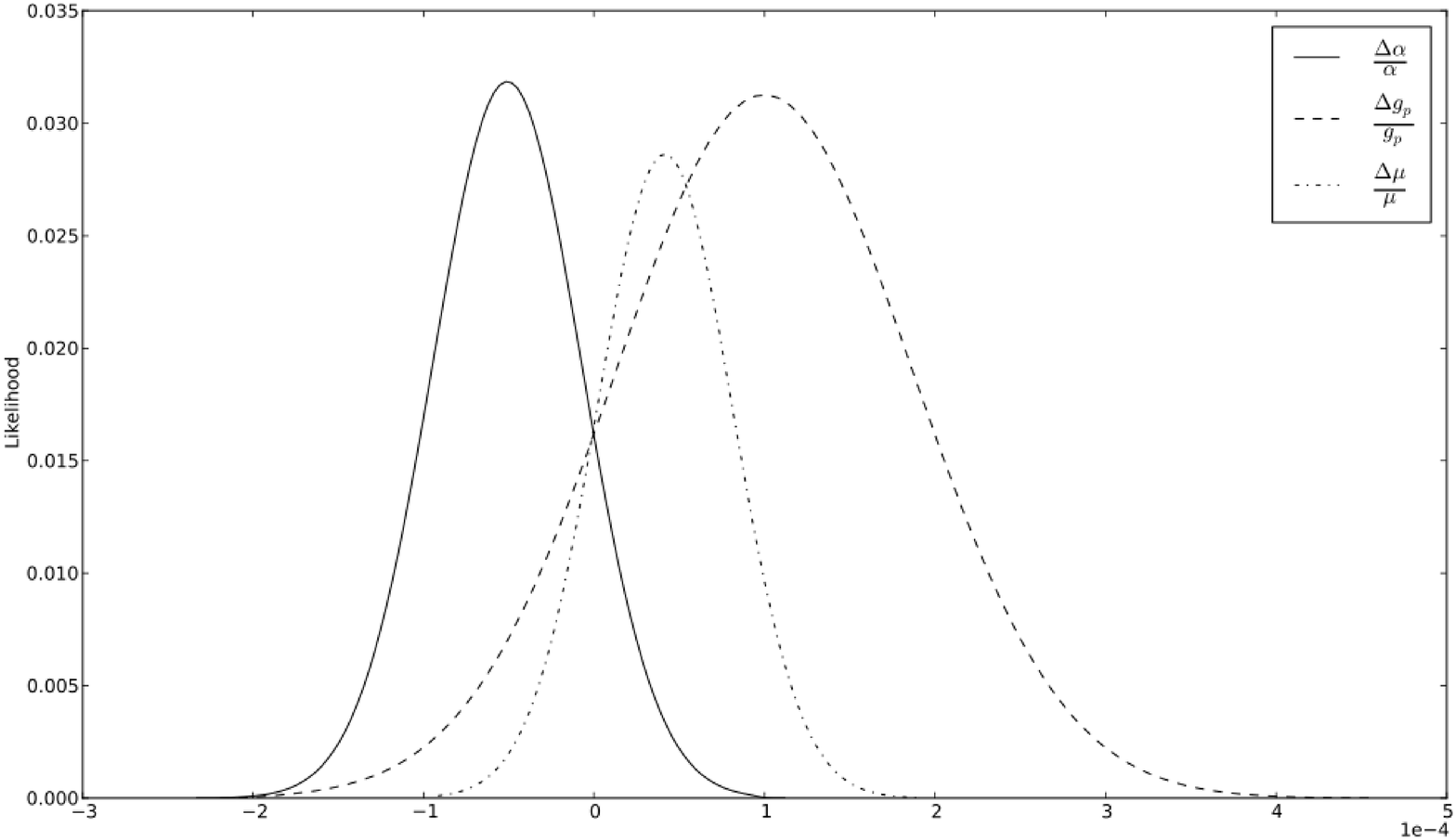}}
\caption{\footnotesize
One-dimensional relative likelihoods (marginalised over the other quantities) for the relative variations of $\alpha$, $\mu$ and $g_{p}$ between $z\approx0.247$ and $z=0$. Reprinted, with permission, from \citet{ferreira2}.
}
\label{qua2}
\end{figure}

From the three measurements we can obtain individual bounds on the variation of each of the couplings. Fig. \ref{qua} shows the two-dimensional likelihood contours in two of the three relevant planes (the third plane is omitted due to the obvious degeneracies between the three parameters). The corresponding one-dimensional relative likelihoods are shown in Fig. \ref{qua2}.

At the one-sigma (68.3\%) confidence level we find
\begin{equation}
\frac{\Delta \alpha}{\alpha} = (-5.1\pm 4.3)10^{-5}
\end{equation}

\begin{equation}
\frac{\Delta \mu}{\mu} = (4.1\pm 3.9)10^{-5}
\end{equation}

\begin{equation}
\frac{\Delta g_{p} }{g_{p}} = (9.9 \pm 8.6)10^{-5}
\end{equation}

\noindent and at the two-sigma level all are consistent with a null result. These constraints are still relatively weak, and in particular do not yet provide an independent test of the dipole results \citep{webb}. However, improvements of one order of magnitude in each of the combined measurements should turn this into a stringent test.

\section{Unification theories and variation of fundamental couplings}

As in  M. Juli\~ao's contribution to these proceedings, we wish to describe phenomenologically a class of models with simultaneous variations of several fundamental couplings, in particular the fine-structure constant $\alpha=e^{2}/ \hbar c$, the proton-to-electron mass ratio $ \mu = m_{p}/m_{e}$ and the proton gyromagnetic ratio $g_{p}$. The simplest way to do this is to relate the various changes to those of a particular dimensionless coupling, typically $\alpha$. Then if $\alpha=\alpha_{0} (1+\delta_{\alpha})$ and

\begin{equation}
\frac{\Delta X}{X} = k_{X} \frac{\Delta \alpha}{\alpha}
\end{equation}

\noindent we have $X = X_{0} (1 + k_{X} \delta_{\alpha})$, and so forth. 

The relations between the couplings will be model-dependent.
We follow a previous analysis \citep{coc,luo}, considering a class of grand unification models in which the weak scale is determined by dimensional transmutation and further assuming that relative variation of all the Yukawa couplings is the same. Finally we assume that the variation of the couplings is driven by a dilaton-type scalar field \citep{campbell}. For our purposes it's natural to assume that particle masses and the QCD scale vary, while the Planck mass is fixed. We then have

\begin{equation}
\frac{\Delta m_{e}}{m_{e}} =  \frac{1}{2} (1 + S) \frac{\Delta \alpha}{\alpha}
\end{equation}

\noindent since the mass of elementary particles is simply the product of the Higgs VEV and the corresponding Yukawa coupling and

\begin{equation}
\frac{\Delta m_{p}}{m_{p}} =  [ 0.8R + 0.2(1 + S)] \frac{\Delta \alpha}{\alpha}.
\end{equation}

The latter equation is the more model-dependent one, as it requires modelling of the proton. At a phenomenological level, the choice $S = -1$, $R = 0$ can also describe the limiting case where $\alpha$ varies but the masses donÕt.
With these assumptions one obtains that the variations of $\mu$ and $\alpha$ are related through

\begin{equation}
\frac{\Delta \mu}{\mu} = [ 0.8R - 0.3(1 + S)] \frac{\Delta \alpha}{\alpha},
\label{dmu}
\end{equation}

\noindent where $R$ and $S$ can be taken as free phenomenological (model-dependent) parameters. Their absolute value can be anything from order unity to several hundreds, although physically one usually expects them to be positive. For the purposes of our analysis they are taken as free parameters to be solely constrained by data.

Further useful relations can be obtained for the proton g-factor \citep{flambaum3,flambaum1,flambaum2},

\begin{equation}
\frac{\Delta g_{p}}{g_{p}} = [ 0.10R - 0.04(1 + S)] \frac{\Delta \alpha}{\alpha}.
\label{dgp}
\end{equation}

Together, these allow us to transform any measurement of a combination of constants into a constraint on the ($\alpha$ R, S) parameter space. Moreover, since $R$ and $S$ are presumed to be universal parameters, once constraints at several redshifts are expressed in this plane the comparison between them becomes unambiguous.

\section{Constraints on unification}

The bounds obtained in the previous section can now be translated, using Eqs. \ref{dmu} and \ref{dgp} into constraints on the phenomenological unification parameters $R$ and $S$. The two functions of $R$ and $S$ are constrained by the existent astrophysical measurements to be

\begin{equation}
0.8R - 0.3(1+S)= -0.81 \pm 0.85,
\end{equation}

\begin{equation}
0.10R - 0.04(1+S) = -1.96 \pm 1.79\,,
\end{equation}

\noindent and solving the previous equations leads to the most likely values of R and S 

\begin{equation}
R \approx 277.8,
\end{equation}

\begin{equation}
S \approx 742.5.
\end{equation}

\begin{figure}[t!]
\resizebox{\hsize}{!}{\includegraphics[width=0.5\textwidth]{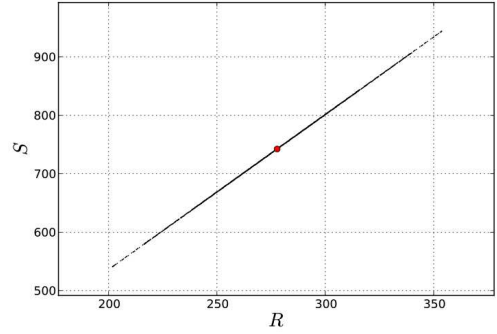}}
\caption{\footnotesize
Combined contours of the two-dimensional likelihood in the $R-S$ plane for the astrophysical measurements towards PKS1413$+$135. Solid, dashed and dotted lines correspond to one-, two- and three-sigma contours (68.3\%, 95.4\% and 99.97\% likelihood, respectively). Reprinted, with permission, from \citet{ferreira2}.
}
\label{res}
\end{figure}

Each of the above relations will determine a degeneracy direction in this plane and the combination of the two is then shown in Fig. \ref{res}.

\begin{figure}[t!]
\resizebox{\hsize}{!}{\includegraphics[width=0.5\textwidth]{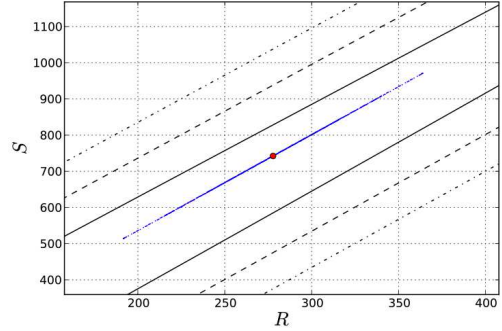}}
\caption{\footnotesize
Two-dimensional likelihood contours in the $R-S$ plane. The broader contours correspond to constrains at $z=0$, coming from atomic clocks \citep{ferreira1}. The smaller contours correspond to the combination of the atomic clock constraints with the ones of PKS1413$+$135. Solid, dashed and dotted lines correspond to one-, two- and three-sigma contours (68.3\%, 95.4\% and 99.97\% likelihood, respectively). Reprinted, with permission, from \citet{ferreira2}.
}
\label{res_comb}
\end{figure}

Finally, fig. \ref{res_comb} shows the likelihood contours for $R$ and $S$ in this case, as well as the result of the combination of the atomic clocks and PKS1413$+$135. The latter results dominate the analysis, and the result is a sub-region from the band in the R-S plane defined by the atomic clock data. From this analysis we can finally obtain the best-fit values for $R$ and $S$; at the one-sigma confidence level we obtain

\begin{equation}
R = 277 \pm 24,
\end{equation}

\begin{equation}
R = 742 \pm 65.
\end{equation}

It has been argued \citep{coc} that typical values for these parameters are $R \approx 30$ and $S \approx 160$ (although these values certainly include a degree of uncertainty). Current constraints from atomic clocks \citep{ferreira1} are fully consistent with these values, but our present analysis shows that this is not the case for PKS1413$+$135.

\section{Conclusions}

We have used available measurements of combinations of dimensionless fundamental couplings towards PKS1413$+$135 in order to obtain individual constraints of the variations of $\alpha$, $\mu$ and $g_{p}$. The precision of the available measurements is not yet sufficient to provide a useful test of the spatial dipole scenario of \citet{webb}. However, if improvements of one order of magnitude or more in each of the combined measurements can be achieved by the next generation of observational facilities (which is a very realistic possibility), this source will eventually provide a powerful test.

We have also used our results to derive constraints on the class of unification scenarios \citep{coc}, and compared them with those previously obtained from local atomic clock measurements. Our analysis shows that both types of measurements prefer unification models characterised by a particular combination of the phenomenological parameters $R$ and $S$, with the PKS1413$+$135 providing stronger constraints on these parameters. It is noteworthy that the parameter values preferred by the current data do not coincide with (arguably naive) expectations on unification scenarios.

We note that in this workshop N. Kanekar presented a yet unpublished revised version of the measurement of \citet{kanekar} (using data from WRST and Arecibo, as opposed to only WRST). This revised measurement provides a weaker detection of a variation, and in that case our constraints discussed above become correspondingly weaker. A re-analysis will be discussed elsewhere.

In any case our results constitute a proof of concept for these methods and motivate the interest of further, more precise measurements of fundamental couplings towards this and other similar astrophysical sources. More generally, they also highlight the point that the early universe is an ideal laboratory in which to carry out precision consistency tests of our standard cosmological paradigm and search for and constrain new physics. Future facilities such as ALMA, the E-ELT, the SKA and others will play a key role in this endeavour.

\begin{acknowledgements}

We acknowledge the financial support of grant PTDC/FIS/111725/2009 from FCT (Portugal). CJM is also supported by an FCT Research Professorship, contract reference IF/00064/2012.

\end{acknowledgements}

\bibliographystyle{aa}

\begin{thebibliography}{}

\bibitem[Campbell et al. (1995)]{campbell}Campbell, B.\,A. {\it et al.}, 1995, Phys. Lett. B345, 429.
\bibitem[Coc et al.(2007)]{coc}Coc, A. {\it et al.}, 2007, Phys. Rev. D76, 023511.
\bibitem[Darling(2004)]{darling}Darling, J., 2004, Astrophys. J. 612, 58.
\bibitem[Ferreira et al. (2012)]{ferreira1}Ferreira, M.\,C. {\it et al.}, 2012, Phys. Rev. D86, 125025.
\bibitem[Ferreira et al. (2013)]{ferreira2}Ferreira, M.\,C. {\it et al.}, 2013, Phys. Lett. B714, 1.
\bibitem[Flambaum (2003)]{flambaum3}Flambaum, V., 2003, arXiv:physics/0302015.
\bibitem[Flambaum et al. (2004),]{flambaum1}Flambaum, V. {\it et al.}, 2004, Phys. Rev. D69, 115006.
\bibitem[Flambaum et al. (2006)]{flambaum2}Flambaum, V. {\it et al.}, 2006, Phys. Rev. C73, 055501.
\bibitem[Kanekar et al.(2010)]{kanekar}Kanekar, N. {\it et al.}, 2010, Astrophys. J. Lett. 716, L23.
\bibitem[Luo et al. (2011)]{luo}Luo, F. {\it et al.}, 2011, Phys. Rev. D84, 096004.
\bibitem[Murphy et al.(2001)]{murphy}Murphy, M. {\it et al.}, 2001, Mon. Not. Roy. Astron. Soc. 327, 1244.
\bibitem[Webb et al. (2011)]{webb}Webb, J. {\it et al.}, 2011, Phys. Rev. Lett. 107, 191101.

\end{thebibliography}

\end{document}